\documentclass[ aps,%
floatfix,%
final,%
notitlepage,%
oneside,%
onecolumn,%
nobibnotes,%
nofootinbib,%
superscriptaddress,%
showpacs,%
centertags]%
{revtex4}

\begin{document}

\title{Determination of $\chi_b$ spin in the decays $\chi_{b0,2}\to\JP\JP$ and $\chi_{b0,2}\to\JP\JP\to 2(\mpmm)$}
\author{ A. V. Luchinsky}
\email{Alexey.Luchinsky@ihep.ru}
\affiliation{Institute for High Energy Physics, Protvino, Russia}
\begin{abstract}
The decays $\chi_{b0,2}\to\JP\JP$ and $\chi_{b0,2}\to\JP\JP\to2(\mpmm)$ are considered. It is shown that with the help of $\JP$-meson angular distributions and the distributions by final muon kinematical variables it is possible to determine the spin of the decaying $\chi_b$-meson. These distributions can also give an additional information about $\JP$-meson wave functions.
\end{abstract}
\maketitle

\section{Introduction}

It is well known that heavy quarkonia \cite{Brambilla:2004wf} give a wonderful possibility to study the strong interactions both at large and at small distances. The processes with these quarkonia split into two stages. The first stage is the creation of quark-antiquark pair $(Q\bar Q)$, while the second one is the subsequent hadronization of this pair into experimentally observed meson. The consequence of the large difference between heavy quark mass $m_Q$ and the strong interaction scale $\Lambda_\mathrm{QCD}$ is that the characteristic distances of these stages are also different. The characteristic distance for the quark-antiquark pair production is $\sim 1/m_Q$, while the scale of its subsequent hadronization is $1/(m_Q v) \gg 1/m_Q$ (here $v\ll 1$ is the typical velocity of the heavy quark in meson). As a result of  this difference the production of quark-antiquark pair and its hadronization are almost independent and heavy quarkonia  make it possible to check quantum chromodynamics both at small ($\sim 1/m_Q$) and large ($\sim (m_Q v)^{-1}$) distances.

Another interesting feature of heavy quarkonia is that the lightest of them lie below the open flavour states production threshold. As a result the decays that are allowed by the Zweig rule (for example $\JP\to D\bar D$ or $\Upsilon(1S) \to B\bar B$), are forbidden by kinematics, so such mesons can decay only via the annihilation of quark-antiquark pair. So the total widths of the ground states of heavy quarkonia are small. One of the consequences of this fact is that the branching fractions of the leptonic decays $\JP,\Upsilon(1S)\to\mpmm$ are large and these decays can be used for the registration of these mesons.

For heavy quarkonia with quantum numbers different from $J^{PC}=1^{--}$ the leptonic decays cannot be used for registration, since the branching fractions of these decays are small. As a result such particles as scalar and tensor bottomonia $\chi_{b0,2}$ are studied rather poor (at the moment even the total widths of these mesons are not known experimentally). On the other hand, it is expected that these mesons should be produced at hadron colliders with large probability. So we need some reliable method of the registration of these particles. In the paper \cite{Kartvelishvili:1984en} it was proposed to use the decays $\chi_{b0,2}\to\JP\JP$ for $\chi_{b0,2}$ detection. In the subsequent analysis \cite{Braguta:2005gw} it was shown, that taking into consideration the relative motion of $c$-quark inside $\JP$-meson one can increase the branching fractions of these decays. This enlargement makes the detection of $\chi_b$-mesons even more promising.

In this article we continue the consideration of this question and study the reaction $\chi_{b0,2}\to\JP\JP$ with the subsequent decays of the $\JP$-meson into leptonic pairs. These decays are interesting by two reasons. First of all, $\JP$-mesons are detected by their leptonic decays, so muons will be the experimentally observed particles in these reaction. The second reason is that from the distributions by final muons one can get an additional information about both $\chi_b$ (for example, one will be able to determine the spin of the bottomonium) and $\JP$-mesons.

The rest of this paper is organized as follows. In the next section we will study the decays $\chi_{b0,2}\to\JP\JP$ and consider the possibility of the determination of $\chi_b$ spin from $\JP$-meson angular distributions. In section \ref{sec:chi4mu} the expressions for the differential widths of the decays $\chi_{b0,2}\to\JP\JP\to 2(\mpmm)$ are presented. In the section \ref{sec:distr} we study the distributions by different kinematical variables of the final muons.

\section{$\chi_{0,2} \to \JP\JP$}

In our paper we will use the helicity amplitude formalism \cite{Haber:1994pe}. According to this method the amplitude of the process $\chi_{b0,2}\to\JP\JP$ is written in the form
\beqa
\mathcal{M}^{(J)}_{\lam_a\lam_b;\mu}(\theta_\chi,\phi_\chi) &=& e^{i\mu\phi_\chi} d^{(J)}_{\mu\lam}(\theta_\chi) \tAJ_{\lam_a,\lam_b},
\eeqa
where $J=0,2$ and $\mu$ are the spin of the initial $\chi_b$-meson and its projection on the fixed axis, $\lam_{a,b}$ are the helicities of $\JP$-mesons, $\lam=\lam_a-\lam_b$, $d^{(J)}_{\mu\lam}$ is the Wigner d-function, and $\theta_\chi$ and $\phi_\chi$ are the polar and azimuthal angles determining the direction of $\JP$ momentum in $\chi$-meson rest frame. After this substitution all dependence on the  kinematics of the process is contained in d-functions and the reduced helicity amplitudes $\tAJ_{\lam_a,\lam_b}$ depend only on the properties of the initial and final mesons.  In the work \cite{Anselmino:1992rw} the explicit expressions for these amplitudes through the wave functions of $\JP$-mesons are presented. Here we would like to mention some general properties of $\tAJ_{\lam_a,\lam_b}$ that will be useful in following. First of all, due to the angular momentum conservation the following inequalities hold
\beqa
|\lam_{a,b}|&\le&1,\qquad |\mu|,|\lam_a-\lam_b| \le J.
\eeqa
Second restriction is caused by the Boze symmetry and leads to the relations
\beq
\tAJ_{\lam_a,\lam_b} &=& \tAJ_{\lam_b,\lam_a} = \tAJ_{-\lam_a,-\lam_b}.
\label{eq:tAJ}
\eeq
These relations decrease the number of independent helicity amplitudes significantly. Namely, we have only 2 independent reduced helicity amplitudes for scalar meson and 4 for tensor one. There is also a chiral suppression
\beq
\tAJ_{\lam_a,\lam_b} &\sim & \left( \frac{m}{M} \right)^{|\lam_a+\lam_b|},
\label{eq:supp}
\eeq
where $m$ and $M$ are the masses of vector charmonium $\JP$ and bottomonium $\chi_b$ respectively.

In \cite{Braguta:2005gw} it was shown, that the numerical values of the reduced helicity amplitudes depend strongly on the choice of the wave functions of charmonium meson. In our paper we will consider some specific cases. Namely, we will use the so-called $\delta$-distribution
\beq
\phi_L(x) &=& \phi_T(x) = \delta\left(x-\frac{1}{2}\right),\label{eq:delta}
\eeq
that corresponds to quarks that are at rest in the $\JP$-meson rest frame, the distribution
\beq
\phi_L(x) &=& 13.2 x(1-x) -36 x^2 (1-x)^2,\label{eq:CZL}\\
\phi_T(x) &=& 30 x^2(1-x^2)\label{eq:CZT},
\eeq
that was proposed in the work \cite{Chernyak:1983ej} for $\rho$-meson, and
\beq
\phi_L(x) &=& \phi_T(x) = 140 x^3(1-x^3),\label{eq:phi3}
\eeq
that was introduced in \cite{Chliapnikov:1977fc}. In what follows the distribution (\ref{eq:delta}) will be labeled as "$\delta$", the set (\ref{eq:CZL},\ref{eq:CZT}) will be labeled as "CZ", and the distribution (\ref{eq:phi3}) as "$\phi_3$". In the paper \cite{Braguta:2005pp} the duality relation that connects the widths of the exclusive decays $\chi_{b0,2}\to\JP\JP$ with the widths of the inclusive decays $\chi_{b0,2}\to\JP D\bar D+X$ was studied. With the help of this relation one can set some restrictions on the branching fraction $\Br(\chi_{b0,2}\to\JP\JP)$. From all of the distributions presented above only the distribution (\ref{eq:phi3}) gives the branching fractions that satisfy these restrictions. In our paper we, nevertheless, will present the results obtained with the use of all the distributions listed above and check whether it is possible to get the information on $\JP$-meson wave functions from the analysis of $\chi_{b}\to\JP\JP$ and $\chi_b\to\JP\JP\to2(\mpmm)$ decays.

In addition to $\JP$-meson wave functions the reduced helicity amplitudes depend also on the values of $\JP$-meson longitudinal and transverse constants $f_{L,T}$ and the derivative of the $\chi_b$-meson wave function at the origin. The longitudinal constant $f_L$ is expressed through the width of $\JP$-meson leptonic decay:
\beqa
\Gamma(\JP\to\epem) &=& \frac{4\pi}{3} e_c^2 \alpha_\mathrm{QED}^2 \frac{f_L^2}{m},
\eeqa
where $\alpha_\mathrm{QED}$ is the fine structure constant and $e_c=2/3$ is the electric charge of the $c$-quark. Hereafter we will suppose that $f_T=f_L$. The derivative of the $\chi_b$-meson wave function is connected with its total widths:
\beqa
\Gamma(\chi_{b0}) &\approx & \Gamma(\chi_{b0}\to gg) = 96\frac{\alpha_s^2}{M^4} |R'(0)|^2,\\ 
\Gamma(\chi_{b2}) &\approx & \Gamma(\chi_{b2}\to gg) = \frac{128}{5}\frac{\alpha_s^2}{M^4} |R'(0)|^2,
\eeqa
where $\alpha_s\approx 0.2$ is the strong coupling constant at the scale of $b$-quark mass. Unfortunately, the values of these widths are not known experimentally, so in what follows we will use the value $|R'(0)|^2=1.34\,\mathrm{GeV}^5$, that is a result of Schrodinger equation solution \cite{Olsson:1984im}. The values of the reduced helicity amplitudes obtained using these numbers are presented in the table \ref{tab:tAJ} (only the independent amplitudes are given in this table, the others can be obtained from them with the help of the relations (\ref{eq:tAJ}) ). 

\begin{table}
$$
\begin{array}{|c||c|c|c||c|c|c|c|c|}
\hline & \tA^{(0)}_{0,0} & \tA^{(0)}_{1,1} & \Gamma_0 & \tA^{(2)}_{0,0} & \tA^{(2)}_{1,1} & \tA^{(2)}_{1,-1} & \tA^{(2)}_{1,0} & \Gamma_2\\
 \hline 
\delta &-0.0054 & 0.0013 & 0.0092 & -0.0066 & -0.0019 & -0.012 & -0.0052 & 0.024 \\ 
 \mathrm{CZ} &-0.021 & 0.0019 & 0.13 & -0.023 & -0.0027 & -0.018 & -0.01 & 0.088 \\ 
 \phi_3 &-0.0075 & 0.0017 & 0.017 & -0.0083 & -0.0024 & -0.015 & -0.0066 & 0.041 \\ 
 \hline
\end{array}
$$
\caption{The reduced helicity amplitudes $\tAJ_{\lam_a,\lam_b}$ (in $\mathrm{GeV}^{1/2}$) and the widths $\Gamma^{(J)}_2$ (in keV)
\label{tab:tAJ}}
\end{table}

\begin{figure}
\begin{center}
\includegraphics{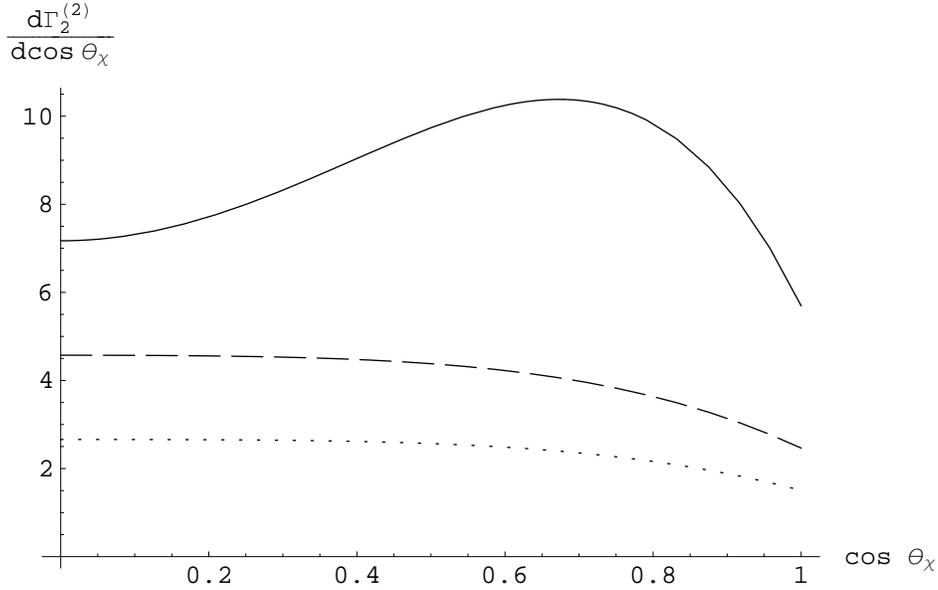}
\caption{
Angular distributions $d\Gamma(\chi_{b2}\to\JP\JP)/d\cos\theta_\chi$ (in eV) versus the cosine of the scattering angle for CZ (solid line), $\phi_3$ (dashed line), and $\delta$ (dotted line) $\JP$-meson wave functions
}\label{fig:dG2dZ}
\end{center}
\end{figure}

Let us consider the decay $\chi_{bJ}\to\JP\JP$. If $\hr_\mu$ is the probability of the $\chi$-meson spin having the projection $\mu$ on the fixed axis, than the width of the decay is equal to
\beqa
\Gamma^{(J)}_2= \frac{1}{4M}
  \sum\limits_{\mu=-J}^J \hr_\mu \sum\limits_{\lam_{a,b}=-1}^1\int\,d\Phi_2^\chi 
  \left| \M^{(J)}_{\lam_a\lam_b;\mu}(\theta_\chi,\phi_\chi)\right|^2,
\eeqa
where
\beqa
d\Phi_2^\chi &=& d\Phi_2(\chi\to\JP\JP)=(2\pi)^4 \delta^4(Q-p_a-p_b) \frac{d^3p_a}{2E_a(2\pi)^3}\frac{d^3p_b}{2E_b(2\pi)^3}
\eeqa
is the lorentz-invariant phase space of the decay $\chi_{bJ}\to\JP\JP$ written in the $\chi_b$ rest frame. From this expression we obtain the $\JP$-meson angular distribution
\beqa
\frac{1}{\Gamma_2^{(J)}}\frac{d\Gamma_2^{(J)}}{d\cos\theta_\chi} &=&
  \frac{1}{\sum\left|\tAJ_{\lam_a\lam_b}\right|^2}
  \sum\limits_\mu \hr_\mu\sum\limits_{\lam_{a,b}}
  	\frac{\left|d^{(J)}_{\mu\lam}(\theta_\chi)\right|^2}{2}\left| \tAJ_{\lam_a,\lam_b}\right|^2.
\eeqa
If we consider $\chi_{b0}$-meson or an unpolarized $\chi_{b2}$-meson (that is with $\hr_\mu=(2J+1)^{-1}$) we get an isotropic distribution. If $\chi_{b2}$-meson is polarized ( this is true for $\chi_{b2}$ produced in $p\bar p$-annihilation), the distribution will be anisotropic. For example, in massless QCD we have \cite{Anselmino:1992rw}
\beqa
\hr_2&=& \hr_{-2}=\hr_0=0, \qquad \hr_1 = \hr_{-1}=\frac{1}{2}.
\eeqa
In  figure \ref{fig:dG2dZ} we show the angular distributions of the $\JP$ meson produced in the decay $\chi_{b2}\to\JP\JP$ for different $c$-quark distributions.

\section{$\chi_{0,2}\to \JP\JP \to 2(\mpmm)$}\label{sec:chi4mu}

In the previous section we have shown, that angular distributions of $\JP$-mesons produced in the decays $\chi_{b}\to\JP\JP$ can be used to separate $\chi_{b0}$- and $\chi_{b2}$-mesons and to determine the reduced chiral amplitudes $\tAJ_{\lam_a,\lam_b}$. In this section we will consider the decays
\beq
\chi_{b0,2} &\to & \JP\JP \to 2(\mpmm)
\label{eq:dec4}
\eeq
and study what information we can get from the distributions of the final particles of these decays. The matrix element of the reaction  (\ref{eq:dec4}) can be written in the form
\beq
\M^{(J)}_\mu &=& 
  \frac{1}{p_a^2-M_\JP^2}  \frac{1}{p_b^2-M_\JP^2}
  \sum\limits_{\lam_{a,b}=-1}^1 \M^{(J)}_{\lam_a\lam_b;\mu}(\theta_\chi,\phi_\chi) 
    \N_{\lam_a}(\theta_a,\phi_a)\N_{\lam_b}(\theta_b,\phi_b),
  \label{eq:M4}
\eeq
where $p_{a,b}$ and $\lam_{a,b}$ are momenta and helicities of the intermediate $\JP$-mesons, $\theta_{a,b}$ and $\phi_{a,b}$ are polar and azimuthal angles determining the direction of muon momentum in the rest frame of the parent $\JP$-meson (see figure \ref{fig:kinematics} for the definition of these angles), and $\N_{\lam}(\theta,\phi)$ is the helicity amplitude of the decay $\JP\to\mpmm$, that depends on the helicity of the vector meson and the muon momentum direction in the $\JP$-rest frame. From the expression (\ref{eq:M4}) we get the width of the decay (\ref{eq:dec4}):
\begin{figure}
\includegraphics{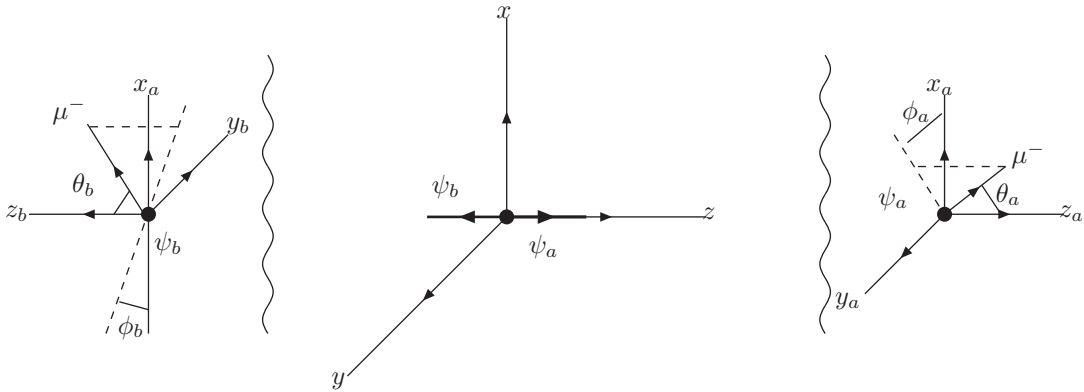}
\caption{Definition of the $\chi_b\to2(\mpmm)$ kinematics. The figure in the center corresponds to $\chi_b$-rest frame, left and right figures correspond to  rest frames of $\JP_b$- and $\JP_a$-mesons.
}\label{fig:kinematics}
\end{figure}
\beq
\Gamma^{(J)}_{4} &\sim&
	\sum\limits_\mu \hr_\mu \int d\Phi_4(\chi\to2(\mpmm)) \left|\M^{(J)}_\mu \right|^2
\sim \nonumber \\ &\sim&
	\sum\limits_\mu \hr_\mu \sum\limits_{\lam_{a,b}} \sum\limits_{\blam_{a,b}}
	\int d\uo d\uo_a d\uo_b \dJ_{\mu\lam}(\theta_\chi)\dJ_{\mu\blam}(\theta_\chi)
\times\nonumber \\ &\times &
	\tAJ_{\lam_a,\lam_b} \tA^{(J)*}_{\blam_a,\blam_b} \U_{\lam_a\blam_a}(\theta_a,\phi_a)\U_{\lam_b\blam_b}(\theta_b,\phi_b).
\label{eq:G4}
\eeq
Here $\blam=\blam_a-\blam_b$, $d\Omega_{a,b}=\sin\theta_{a,b} d\theta_{a,b} d\phi_{a,b}$ are the solid angles or the decays $\JP_{a,b}\to\mpmm$ in $\JP$-meson rest frame, and $\mathcal{U}_{\lam\blam}(\theta,\phi)=\N_\lam(\theta,\phi) \mathcal{N}^*_{\blam}(\theta,\phi)$. The functions $\mathcal{U}_{\lam,\blam}(\theta,\phi)$ satisfy the relations
\beqa
\mathcal{U}_{\lam,\blam}(\theta,\phi) &=& \mathcal{U}^*_{\blam,\lam}(\theta,\phi)
  = (-1)^{\lam-\blam} \mathcal{U}_{-\blam,-\lam}(\theta,\phi),
\eeqa
and the explicit expression are
\beqa
\mathcal{U}_{1,1}(\theta,\phi) &=& m^2g^2(2-\sin^2\theta) =\mathcal{U}_{-1,-1}(\theta,\phi)\\
\mathcal{U}_{0,0}(\theta,\phi) &=& 2m^2g^2\sin^2\theta \\
\mathcal{U}_{1,0}(\theta,\phi) &=& \sqrt{2}m^2g^2 \sin\theta\cos\theta e^{i\phi} =-\mathcal{U}_{0,-1}(\theta,\phi)\\
\mathcal{U}_{0,1}(\theta,\phi) &=& \sqrt{2}m^2g^2 \sin\theta\cos\theta e^{-i\phi} =-\mathcal{U}_{-1,0}(\theta,\phi)\\
\mathcal{U}_{1,-1}(\theta,\phi) &=& m^2g^2 \sin^2\theta e^{2i\phi}.
\eeqa
Here $g$ is the effective constant of the $\JP\mpmm$ vertex, that can be expressed through the width of $\JP$ leptonic decay:
\beqa
g^2 &=& 12\pi\frac{\Gamma(\psi\to\mu^-\mu^+)}{m}
\eeqa
and we have neglected the muon mass. When deriving the formula (\ref{eq:G4}) we have used the factorization of the 4-particle phase space
\beqa
d\Phi_4(\chi\to\JP\JP\to\mpmm\mpmm) &\sim& dp_a^2 dp_b^2 d\Phi_2(\chi\to\JP\JP) d\Phi_2(\JP_a\to\mpmm) d\Phi_2(\JP_b\to\mpmm)
\eeqa
and the relation
\beqa
\left|\frac{1}{p_{a,b}^2-M_\JP^2 }\right|^2 &\sim& \frac{1}{\M_\JP\Gamma_\JP} \delta\left(p_{a,b}^2-M_\JP^2 \right),
\eeqa
that is caused by the smallness of the $\JP$-meson total width.

\section{Final muon distributions\label{sec:distr}}

From the formula (\ref{eq:G4}) we can derive the distributions by the kinematic variables of final muons. These distributions will be considered in this section.

\begin{figure}
\includegraphics{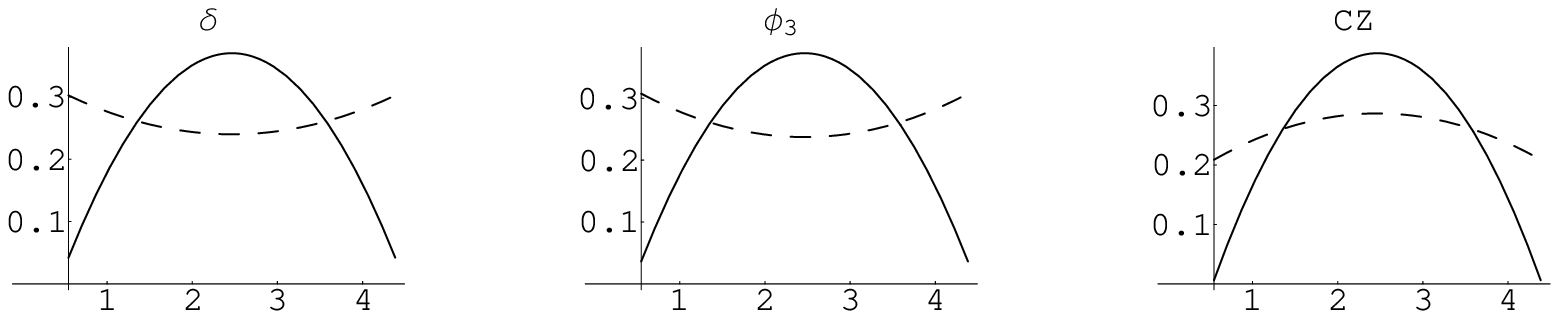}
\caption{Normalized muon energy distributions in the decays $\chi_{b0}\to 2(\mpmm)$ (solid line) and $\chi_{b2}\to2(\mpmm)$ (dashed line) for different sets of $\JP$-meson wave functions}\label{fig:dGdEm}
\end{figure}

Let us consider the case when only one $\mpmm$-pair is registered. In this case we can construct a normalized helicity density matrix
\beqa
\rho_{\lam_a,\blam_a} &= &
 \frac{\sum\limits_{\lam_b}\tAJ_{\lam_a,\lam_b}\tA^{(J)*}_{\blam_a,\lam_b}}{\sum\limits_{\lam_a,b}|\tAJ_{\lam_a,\lam_b}|^2}.
\eeqa
The distribution by the angle $\theta_a$ has the form
\beqa
\frac{1}{\Gamma}\frac{d\Gamma}{d\cos\theta_a}  &=&
  \frac{3}{4}\left\{
    (1-\rho_{0,0})\cos^2\theta_a +\frac{1+\rho_{0,0}}{2}\sin^2\theta_a
  \right\}.
\eeqa
Using this expression one can construct the distributions by the experimentally observable variables.
 If $E_\mu$ is the energy of the final muon in the rest frame of the initial bottomonium, than the distribution by this variable has the form
\beq
\frac{1}{\Gamma}\frac{d\Gamma}{d E_\mu} &=& \frac{3}{2M\beta}\left\{
  1+\rho_{0,0} -\frac{3\rho_{0,0}-1}{\beta^2} \left( \frac{4E_\mu}{M}-1\right)^2
\right\},
\label{eq:dGdth}
\eeq
where $\beta=\sqrt{1-4m^2/M^2}$ is the $\JP$ velocity in the $\chi_b$ rest frame. It is clearly seen that the form of this distribution depends strongly on the value of the parameter $\rho_{0,0}$. If $\rho_{0,0}=1/3$ the differential width will not depend on the energy of the muon, whereas for $\rho_{0,0}>1/3$ or $\rho_{0,0}<1/3$ we will have a parabola with positive or negative main coefficient. The parameter $\rho_{0,0}$ is expressed through the reduced helicity amplitudes according to
\beqa
\rho_{0,0} &=& \frac{ |\tA_{0,0}|^2+2|\tA_{1,0}|^2}{|\tA_{0,0}|^2+2|\tA_{1,-1}|^2+2|\tA_{1,1}|^2+4|\tA_{1,0}|^2}.
\eeqa
In the second section we have mentioned that the reduced amplitudes $\AJ_{\lam_a,\lam_b}$ should satisfy the restrictions $|\lam_a-\lam_b|\le J$ and the chiral suppression (\ref{eq:supp}) should take place. A result of these restrictions is that in the massless $\JP$ limit we have $\rho_{0,0}=1$ for scalar bottomonium and $\rho_{0,0}=1/3$ for tensor one. Since the mass of the $\JP$-meson is small compared with the mass of $\chi_b$, similar results should take place also for real $\JP$, so the forms of the muon energy distributions for scalar and tensor bottomonia a different. In figure \ref{fig:dGdEm} we show the distribution $d\Gamma/dE_\mu$ for scalar and tensor bottomonia (solid and dashed lines respectively) for different sets of the $\JP$-meson wave functions. It is clearly seen that from this distribution one can determine the spin of $\chi_{b}$ meson and set the restrictions on the $\JP$-meson wave function.
Similar situation is observed also for the distributions by the transverse muon momentum $k_\perp$, that is the component of the muon momentum that is orthogonal to the momentum of the parent $\JP$-meson. From the formula (\ref{eq:dGdth}) one can easily obtain the analytical expression for this distribution:
\beqa
\frac{1}{\Gamma}\frac{d\Gamma}{dk_\perp} &=& \frac{3}{m}\frac{k_\perp}{\sqrt{m^2-4k_\perp^2}} \left\{
  1-\rho_{0,0}-(1-3\rho_{0,0}) \frac{2k_\perp^2}{m_2}
\right\}.
\eeqa
Unfortunately, the first factor masks the form of this distribution, so it is hard to distinguish the scalar bottomonium from the tensor one from it. 

\begin{figure}
\includegraphics{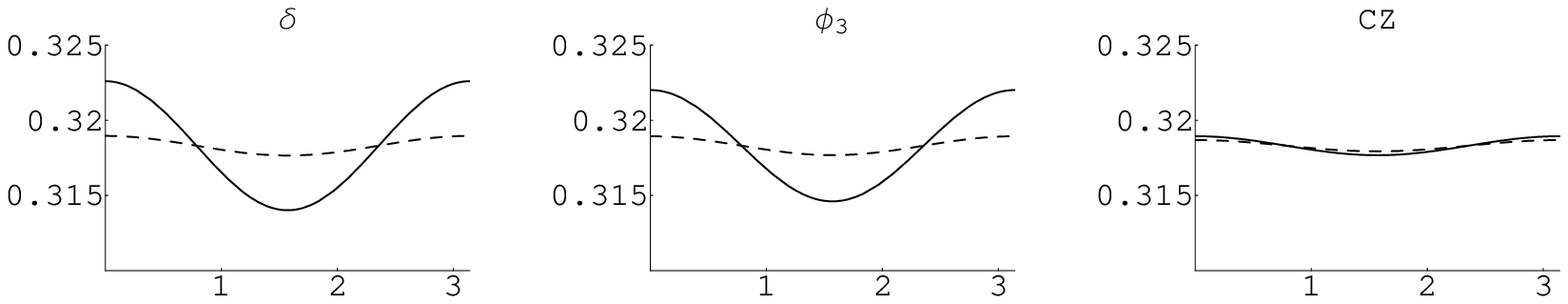}
\caption{
Normalized $\delta\phi$ distributions in the decays $\chi_{b0}\to2(\mpmm)$ (solid line) and $\chi_{b0}\to2(\mpmm)$ (dashed line) for different sets of $\JP$-meson wave functions
}\label{fig:deltaPhi}
\end{figure}

Let us now proceed to the case when the decays of both $\JP$-mesons are  registered. Here we will consider the distribution by the angle between decay planes, that is the planes formed by the final muon momenta. This angle is expressed through the azimuthal angles $\phi_{a,b}$ by the relation $\delta\phi=\phi_a+\phi_b$, and the normalized distribution over this angle have the form
\beqa
\frac{1}{\Gamma_4^{(J)}} \frac{d\Gamma_4^{(J)}}{d\delta\phi} &=&
  \frac{1}{\pi}\left\{
  1+\frac{|\tAJ_{1,1}|^2}{\sum\limits_{\lam_{a,b}}|\tAJ_{\lam_a,\lam_b}|^2} \frac{\cos 2\delta\phi}{4}
  \right\}.
\eeqa
We show the distributions by this angle in the figure \ref{fig:deltaPhi}. It is hard to use them for separating $\chi_{b0}$ and $\chi_{b2}$-mesons, but these distributions give us the value of the matrix element $|\tAJ_{1,1}|$.

\section{Conclusion}

In the works \cite{Kartvelishvili:1984en,Braguta:2005gw} it was proposed to use the decays $\chi_{b0,2}\to\JP\JP$ for the registration of $\chi_{b0,2}$ mesons. Here we continue the consideration of this question and study the decays $\chi_{b0,2}\to\JP\JP\to2(\mpmm)$.

One of the questions that arise in the observation of scalar and tensor bottomonia is the determination of the spin of this particle. Since the difference between their masses is tiny ($M_{\chi_{b2}}-M_{\chi_{b0}}\approx 50$ MeV), it could be smaller than the instrumental error of the detector. That is why it will be rather difficult to determine the spin of the meson from its mass and one needs to have some other methods for the separation of $\chi_{b0}$- and $\chi_{b2}$-mesons. In this paper we show, that this separation can be performed using the angular distributions of $\JP$-mesons produced in the reactions $\chi_{b0,2}\to\JP\JP$, or the distributions of the final muons produced in the decays $\chi_{b0,2}\to\JP\JP\to 2(\mpmm)$.

From this distributions one can also get an additional information about the wave functions of intermediate $\JP$-mesons. In this work we consider several sets of these wave functions and show that the the distributions by final muon energy in the $\chi_b$ rest frames for these sets differ from each other. 

The author thanks A.K. Likhoded  for useful discussions. This work was partially
supported by Russian Foundation for Basic Research under grant no.04-02-17530, Russian Education
Ministry grant no.E02-31-96, CRDF grant no.MO-011-0, Scientific School grant no.SS-1303.2003.2.

\end{document}